\documentclass[runningheads]{llncs}
\usepackage{graphicx}
\usepackage{multirow}
\usepackage{tabularx}
\usepackage{hyperref}
\usepackage{amsmath}
\usepackage{multirow}
\usepackage[table,xcdraw]{xcolor}
\usepackage{bbm}
\usepackage{bm}
\usepackage{mathtools}
\usepackage{amssymb}

\DeclarePairedDelimiter{\norm}{\lVert}{\rVert} 
\begin{document}
%
%\title{Contribution Title\thanks{Supported by organization x.}}
\title{BMD-GAN: Bone mineral density estimation using x-ray image decomposition into projections of bone-segmented quantitative computed tomography using hierarchical learning}
\titlerunning{BMD-GAN}
\newcommand{\repeatthanks}{\textsuperscript{\thefootnote}}
\author{****************************************** ************************\inst{1}, ********** ******************\inst{2}}
\institute{********************************************************************** \and ************************************************** \\
\email{****************}}

\author{Yi Gu\inst{1}, Yoshito Otake\inst{1}, Keisuke Uemura\inst{2}, Mazen Soufi\inst{1},\newline Masaki Takao\inst{3}, Nobuhiko Sugano\inst{4}, and Yoshinobu Sato\inst{1}}
% index{Gu, Yi}
% index{Otake, Yoshito}
% index{Uemura, Keisuke}
% index{Soufi, Mazen}
% index{Takao, Masaki}
% index{Sugano, Nobuhiko}
% index{Sato, Yoshinobu}

%
\authorrunning{Yi Gu et al.}
\institute{Division of Information Science, Graduate School of Science and Technology,\newline Nara Institute of Science and Technology \\
\email{\{gu.yi.gu4,otake,yoshi\}@is.naist.jp}
\and Department of Orthopaedics, Osaka University Graduate School of Medicine \\
\and Department of Bone and Joint Surgery,\newline Ehime University Graduate School of Medicine \\
\and Department of Orthopaedic Medical Engineering,\newline Osaka University Graduate School of Medicine \\
}
\maketitle              % typeset the header of the contribution
\begin{abstract}
We propose a method for estimating the bone mineral density (BMD) from a plain x-ray image. Dual-energy X-ray absorptiometry (DXA) and quantitative computed tomography (QCT) provide high accuracy in diagnosing osteoporosis; however, these modalities require special equipment and scan protocols. Measuring BMD from an x-ray image provides an opportunistic screening, which is potentially useful for early diagnosis. The previous methods that directly learn the relationship between x-ray images and BMD require a large training dataset to achieve high accuracy because of large intensity variations in the x-ray images. Therefore, we propose an approach using the QCT for training a generative adversarial network (GAN) and decomposing an x-ray image into a projection of bone-segmented QCT. The proposed hierarchical learning improved the robustness and accuracy of quantitatively decomposing a small-area target. The evaluation of 200 patients with osteoarthritis using the proposed method, which we named BMD-GAN, demonstrated a Pearson correlation coefficient of 0.888 between the predicted and ground truth DXA-measured BMD. Besides not requiring a large-scale training database, another advantage of our method is its extensibility to other anatomical areas, such as the vertebrae and rib bones.
\end{abstract}

\section{Introduction}
The measurement of bone mineral density (BMD) is essential for diagnosing osteoporosis. Although dual-energy X-ray absorptiometry (DXA) \cite{Blake2007} and quantitative computed tomography (QCT) \cite{Mueller2011,Aggarwal2021} are regarded as the gold standards for BMD measurement, there is a strong demand of developing simpler methods for measuring BMD, which provide opportunistic screening for the early detection of osteoporosis in patients without symptoms. To realize this, recent studies have focused on using deep learning to estimate BMD or diagnose osteoporosis from an x-ray image, which is a more widespread modality than DXA and QCT. These studies performed regression to estimate BMD or classification to diagnose osteoporosis, grading directly from x-ray images \cite{Hsieh2021,Ho2021,Jang2021,Yamamoto2020}, some of which achieved high correlations with DXA-measured BMD, and grading using a large-scale training dataset. However, these methods do not provide the spatial density distribution of the target bone and do not leverage information from QCT. Furthermore, the critical requirement of a large-scale training database would limit the application of these methods when the target bone of BMD measurement extends to other anatomical areas (e.g., different positions of vertebrae, pelvis, sacrum, etc.).

From the viewpoint of processing x-ray images of bones, bone suppression is one of the main topics \cite{Yang2017}, enhancing the visibility of other soft tissues and increasing the diagnosis rate by machines and clinicians. These studies used a convolutional neural network, particularly a generative adversarial network (GAN) \cite{Isola2017}, to decompose x-ray images into bones and other soft tissues \cite{Liu2019,Yang2017,Eslami2020}. Because these studies focued on soft tissues, they did not address the quantitative evaluation of bone decomposition for BMD estimation. Despite the difficulty in GAN training, recent studies \cite{zhao2020,Zhang2020,Wu2021} were able to stabilize the training and reduce the demand for a large-scale database. While those studies inspired this study, we introduced the hierarchical learning (HL) method so that the small-area region of the target bone required for BMD estimation is decomposed accurately and stably even without requiring a large training dataset. Unlike the previous methods for BMD estimation from x-ray images \cite{Hsieh2021,Ho2021,Jang2021,Yamamoto2020}, our method fully uses rich information from QCT in its training phase to estimate the density distributions in addition to BMD, that is, the average density within a specific clinically- defined region- of- interest (ROI). Furthermore, it can be applied to any ROI of any bone. In this study, we showed accuracy validations for the density distributions and BMDs estimated from x-ray images by comparing them with BMDs measured by DXA (hereafter “DXA-BMD”), QCT (hereafter “QCT-BMD”), and the average intensity of the ground truth 2D projections of bone-segmented QCT, respectively.

\begin{figure}
    \centering
    \includegraphics[width=\textwidth]{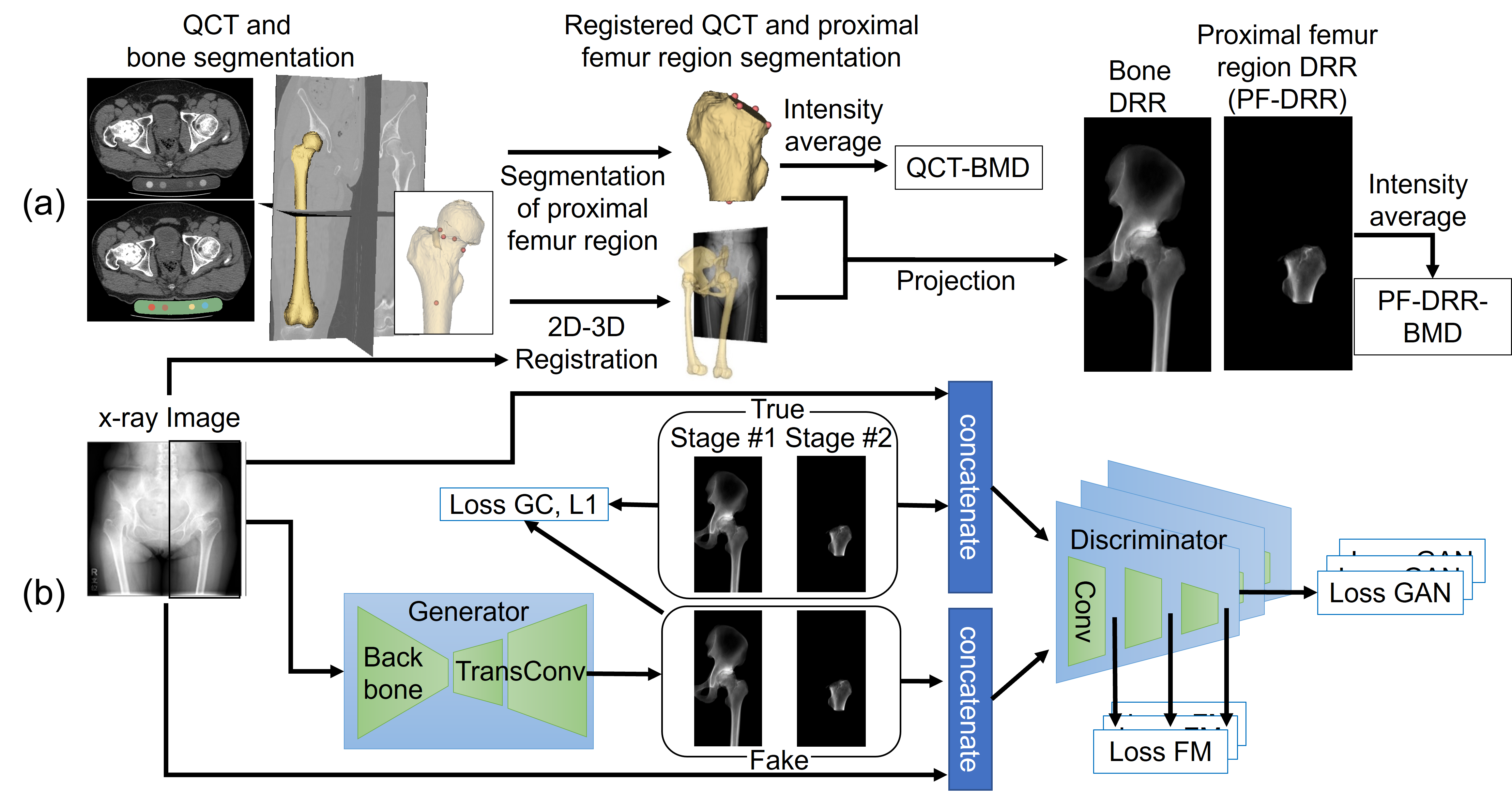}
    \caption{Overview of the proposed method. (a) Construction of the training dataset consisting of the intensity calibration of CT \cite{Uemura2021}, bone segmentation \cite{Hiasa2020}, 2D-3D registration to the x-ray image \cite{Otake2012} and DRR generation by projecting QCT. (b) The network architecture of the proposed BMD-GAN.}
    \label{fig:method}
\end{figure}

\section{Method}
\subsection{Overview of the proposed method}

\begin{figure}
    \centering
    \includegraphics[width=\textwidth]{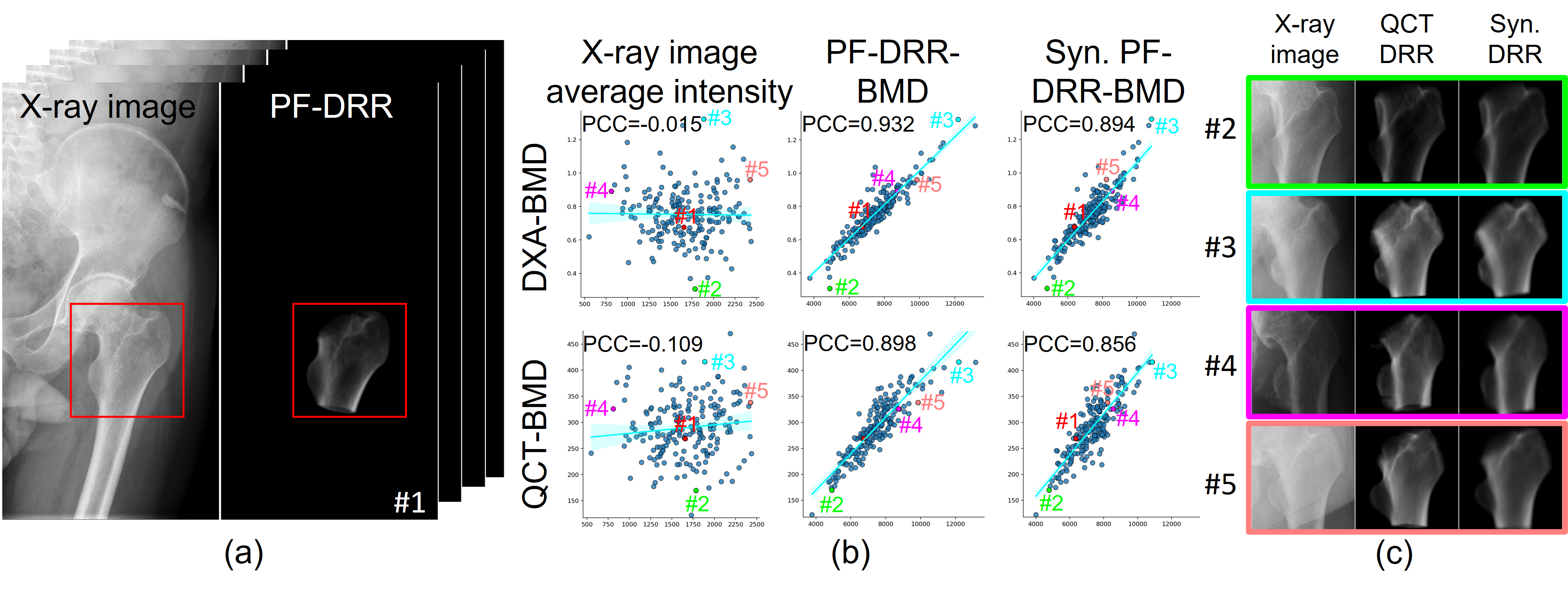}
    \caption{Relationship between the intensities of the x-ray image, DRR, and BMD values in 200 patient datasets. (a) Paired (registered) dataset of x-ray image and DRR; (b) scatter plots showing the correlation of the average intensity of x-ray images, QCT DRR, and synthesized DRR with DXA-measured BMD and QCT-measured BMD; (c) proximal femur ROIs of four representative cases. ROIs \#2 and \#3 have similar x-ray intensity but significantly different BMD, whereas ROIs \#4 and \#5 have similar BMD but significantly different x-ray intensity. The synthesized DRRs correctly recovered the intensity of QCT DRR, regardless of the intensity of the input x-ray image.}
    \label{fig1}
\end{figure}

Figure \ref{fig:method} shows the overview of the proposed method. An image synthesis model decomposes the x-ray image into the digitally reconstructed radiograph (DRR) of the proximal femur region [hereafter “proximal femur region DRR (PF-DRR)”], whose average intensity provides the predicted BMD. Our BMD-GAN applies a hierarchical framework during the training in which the model is first trained to extract the pelvis and femur bones and then the proximal femur region in the subsequent stage. Figure \ref{fig1} illustrates the relationship between x-ray images and BMDs in our patient dataset, demonstrating the challenge in the task of BMD prediction based on an x-ray image.

\subsection{Dataset construction}

In this study, we constructed two datasets: 1) the stage-one dataset containing x-ray images; QCT; the 3D segmentation masks of the pelvis and femur, which were obtained by applying Bayesian U-net \cite{Hiasa2020}; and bone DRR, which were created from QCT using 2D-3D registration \cite{Otake2012} followed by projection with the 3D mask, and 2) the stage-two dataset containing x-ray images; QCT; the 3D mask of the proximal femur region, which is obtained by manually labeled bony landmarks defined in \cite{Uemura2022} by an expert clinician; and the PF-DRR. The construction procedure of the stage-two dataset followed \cite{Uemura2022}. The intensity-based 2D-3D registration using gradient correlation similarity metric and the CMA-ES optimizer \cite{Otake2012} was performed on each patient’s x-ray image and QCT. All x-ray images and DRRs were normalized into the size of 256 $\times$ 512 by central cropping and resizing. The aspect ratio of the original x-ray images varied from 0.880 to 1.228 (width/height). We first split them horizontally in half at the center. Then the side with the target hip was reshaped to a predefined image size (256 $\times$ 512 in this experiment) by aligning the center of the image and cropping the region outside the image after resizing to fit the shorter edge of the width and height.

\subsection{Paired image translation from an x-ray image to a PF-DRR}
The GAN with conditional discriminators was used to train the decomposition model. We followed most settings used in Pix2PixHD \cite{Wang2018} including the multi-scale discriminators and the Feature Matching loss (Loss FM in Fig. \ref{fig:method}), among others. Instead of the ResNet Generator used in Pix2Pix \cite{Isola2017} and Pix2PixHD \cite{Wang2018}, we adopted the state-of-the-art model HRFormer \cite{Yuan2021}, which is a transformer-based model for segmentation to be the backbone of the generator, namely, HRFormer Generator. Instead of using the hierarchical structure of the generator used in Pix2PixHD \cite{Wang2018}, we applied the HL framework in which a two-stage training is used. In the first stage, the model is trained to decompose an x-ray image into the pelvis and femur bones; in the second stage, the target is transferred to the proximal femur region. In the training, we used the adversarial loss $\mathcal{L}_{\mathrm{GAN}}$, which is defined as 
\begin{equation}
    \mathcal{L}_{\mathrm{GAN}}(G,D)=\mathbbm{E}_{(\bm{\mathrm{I^{Xp}}},\bm{\mathrm{I^{DRR}}})}[\log D(\bm{\mathrm{I^{Xp}}},\bm{\mathrm{I^{DRR}}})]+\\
    \mathbbm{E}_{\bm{\mathrm{I^{Xp}}}}[\log (1-D(\bm{\mathrm{I^{Xp}}},G(\bm{\mathrm{I^{Xp}}}))],
    \label{eq:LGAN}
\end{equation}
where $G$, $D$, $\bm{\mathrm{I^{Xp}}}$, and $\bm{\mathrm{I^{DRR}}}$ are the generator, discriminator, x-ray image, and decomposed DRR, respectively. We denote $\mathbbm{E}_{\bm{\mathrm{I^{Xp}}}}\triangleq\mathbbm{E}_{\bm{\mathrm{I^{Xp}}}\sim p_{\mathrm{data}}(\bm{\mathrm{I^{Xp}}})}$ and $\mathbbm{E}_{(\bm{\mathrm{I^{Xp}}},\bm{\mathrm{I^{DRR}}})}\triangleq\mathbbm{E}_{\bm{\mathrm{(I^{Xp}}},\bm{\mathrm{I^{DRR})}}\sim p_{\mathrm{data}}(\bm{\mathrm{I^{Xp}}},\bm{\mathrm{I^{DRR}}})}$. Furthermore, we used the Feature Matching loss $\mathcal{L}_{\mathrm{FM}}$ proposed in \cite{Wang2018}, given by
\begin{equation}
    \mathcal{L}_{\mathrm{FM}}(G,D)=\mathbbm{E}_{(\bm{\mathrm{I^{Xp}}},\bm{\mathrm{I^{DRR}}})}\sum_{i=1}^{T} \\
    \frac{1}{N_{i}}[\norm{D^{(i)}(\bm{\mathrm{I^{Xp}}},\bm{\mathrm{I^{DRR}}})-D^{(i)}(\bm{\mathrm{I^{Xp}}},G(\bm{\mathrm{I^{Xp}}}))}_{1}],
    \label{eq:LFM}
\end{equation}
where $D^{(i)}$, $T$, and $N_{i}$ denote the $i$th-layer feature extractor of discriminator $D$, the total number of layers, and the number of elements in each layer, respectively. We did not use a perceptual loss because a well-pretrained perceptual model is difficult to obtain under a limited dataset. We instead used a simple $L_{1}$ loss $\mathcal{L}_{L1}$ defined as 
\begin{equation}
    \mathcal{L}_{L1}(G)=\mathbbm{E}_{(\bm{\mathrm{I^{Xp}}},\bm{\mathrm{I^{DRR}}})}[\norm{\bm{\mathrm{I^{DRR}}}-G(\bm{\mathrm{I^{Xp}}})}_{1}].
    \label{eq:Ll1}
\end{equation}
To maintain the consistency of the structure between the fake DRR $G(\bm{\mathrm{I^{Xp}}})$ and the true DRR $\bm{\mathrm{I^{DRR}}}$, we regularized the generator with the gradient-matching constraints proposed in \cite{Penney1998}, using the gradient correlation loss $\mathcal{L}_{GC}$ defined as 
\begin{equation}
    \mathcal{L}_{GC}(G)=\mathbbm{E}_{(\bm{\mathrm{I^{Xp}}},\bm{\mathrm{I^{DRR}}})}[NCC(\nabla_{x}\bm{\mathrm{I^{DRR}}},\nabla_{x}G(\bm{\mathrm{I^{Xp}}}))+NCC(\nabla_{y}\bm{\mathrm{I^{DRR}}},\nabla_{y}G(\bm{\mathrm{I^{Xp}}}))],
    \label{eq:LGC}
\end{equation}
where $NCC(\bm{\mathrm{A}},\bm{\mathrm{B}})$ is the normalized cross-correlation of $\bm{\mathrm{A}}$ and $\bm{\mathrm{B}}$, and $\nabla_{x}$ and $\nabla_{y}$ are the $x$ and $y$ components of the gradient vector, respectively. Thus, our full objective was defined as 
\begin{equation}
\begin{split}
    \min_G \left(\lambda_{L1}\mathcal{L}_{L1}(G)+\lambda_{GC}\mathcal{L}_{GC}(G)+\lambda_{FM}\sum_{k=1,2,3}\mathcal{L}_{\mathrm{FM}}(G,D_{k}) \right. \\ \left. +\left(\max_{D_1,D_2,D_3}\sum_{k=1,2,3}\mathcal{L}_{\mathrm{GAN}}(G,D_{k})\right)\right),
    \label{eq:Lall}
\end{split}
\end{equation}
where the multi-scale discriminators $D_1$, $D_2$, and $D_3$ were used under three resolutions as in \cite{Wang2018}, and $\lambda_{L1}$, $\lambda_{GC}$, and $\lambda_{FM}$ are the hyper-parameters that balance the importance of the terms. Both stage-one and stage-two training use the same loss functions. For the learning rate policy, we used the linear decay strategy used in Pix2Pix \cite{Isola2017} in stage-one training and the stochastic gradient descent with warm restarts (SGDR) proposed in \cite{Loshchilov2017} in stage-two training.

Next, the average intensity of the predicted PF-DRR was calculated. Note that the pixels with an intensity equal to or larger than the threshold $t$ were averaged in this study. We empirically defined $t=1000$ in the experiment. The PF-DRR-average of all training datasets was linearly fitted to DXA-BMD and QCT-BMD to obtain the slope and intercept, which were used to convert the PF-DRR-average to the BMDs of the test dataset. 

\subsection{Generator backbone}
In the experiments, we compared the performance of two more models that were used for semantic segmentation as the backbone of generator--DAFormer \cite{Hoyer2021} and HRNetV2 \cite{Wang2020}, namely, DAFormer Generator and HRNet Generator, respectively. Though these state-of-the-art models have been proven to have high performance in segmentation tasks, their ability to decompose images has not been thoroughly assessed. We tuned each backbone’s learning rate, optimizer, weight decay, and epochs, including ResNet, DAFormer, HRNet, and HRFormer, separately.

\begin{figure}
    \centering
    \includegraphics[width=\textwidth]{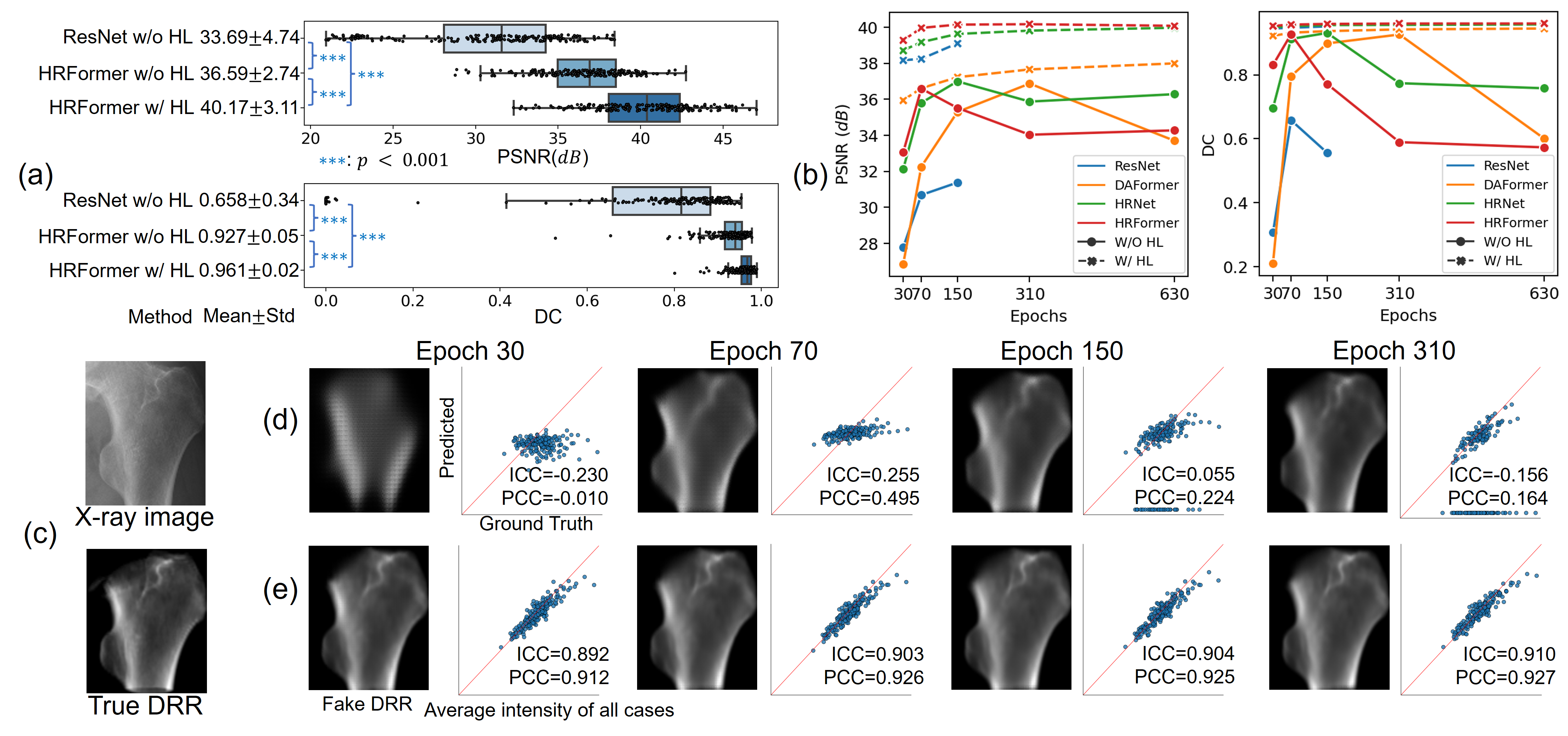}
    \caption{Results of x-ray image decomposition. (a) Evaluation of image decomposition for ResNet (baseline) without HL, HRFormer without HL, and HRFormer with HL. (b) Convergence analysis of each backbone without and with HL. (c) The ROI of the input x-ray image and true DRR, and comparison of the training progress between (d) without HL and (e) with HL.}
    \label{fig:epoch}
\end{figure}

\section{Experiments and Results}
\subsection{Experimental materials, setting, and evaluation metrics}
Ethical approval was obtained from the Institutional Review Boards (IRBs) of the institutions participating in this study (IRB approval numbers: 21115 at Osaka University Hospital and 2021-M-11 at Nara Institute of Science and Technology). The constructed stage-one dataset contained 275 cases. Each case had an x-ray image, and its paired bone DRRs of the left and right sides were split by the vertical middle line, resulting in 525 image pairs after excluding the images with the hip implant. The constructed stage-two dataset contained 200 cases obtained retrospectively from 200 patients (166 females) who underwent primary total hip arthroplasty between May 2011 and December 2015.Each case has an x-ray image and its paired PF-DRR of one side with its ground truth DXA-BMD. The patients’ age and T-scores calculated from the DXA-BMD of the proximal femur were 59.5 $\pm$ 12.9 years (range: 26 to 86) and -1.23 $\pm$ 1.55 (range: -5.68 to 4.47), respectively. The calibration phantom (B-MAS200, Kyoto Kagaku, Kyoto, Japan) \cite{Uemura2021}, which is used to convert radiodensity [in Housunsfield units (HU)] to bone density (in $\mathrm{mg/cm^3}$), contains known densities of hydroxyapatite $\mathrm{Ca_{10}(PO_{4})_{6}(OH)_{2}}$. All CT images used in this study were obtained by the OptimaCT660 scanner (GE Healthcare Japan, Tokyo, Japan); all x-ray images, which were scanned by the devices from FUJIFILM Corporation and Philips Medical Systems, were acquired in the standing position in the anterior-posterior direction; and all DXA images of the proximal femur were acquired for the operative side (Discovery A, Hologic Japan, Tokyo, Japan) to obtain the ground truth DXA-BMD. The evaluations in the following sections were performed on the stage-two dataset. Five-fold cross-validation was performed to investigate the effect of HL and compare the backbones. The ResNet Generator without HL was set as the baseline for evaluating the decomposition accuracy. Furthermore, We compared our best method with the conventional method proposed in \cite{Hsieh2021}, which directly regresses the BMD from the x-ray images, under our limited dataset. 

To evaluate the performance on image decomposition, we used the peak signal-to-noise ratio (PSNR), multi-threshold dice coefficient (DC), intraclass correlation coefficient (ICC), and Pearson correlation coefficient (PCC) of the average intensity of the PF-DRR. To evaluate BMD estimation, we used ICC, PCC, mean absolute error (MAE), and standard error of estimate (SEE). Statistical significance was evaluated using the single-factor repeated measures analysis of variance model. P-values of less than were used to denote statistical significance.

\begin{figure}
    \centering
    \includegraphics[width=\textwidth]{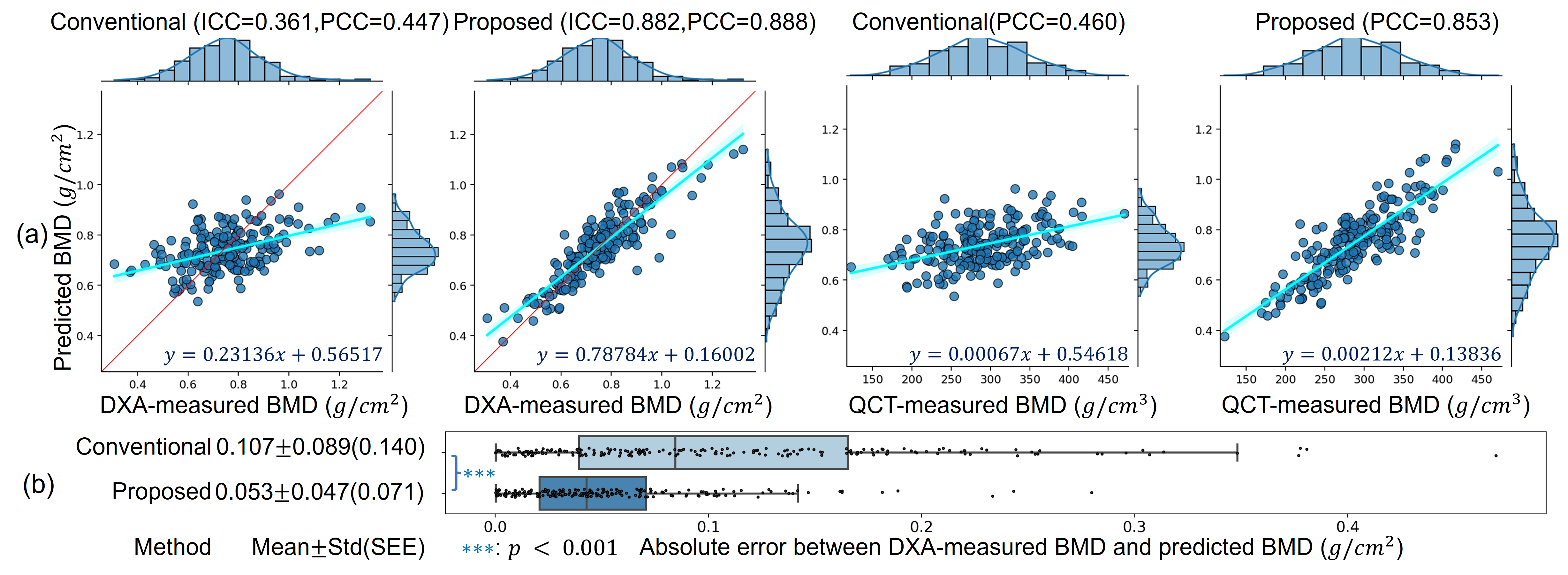}
    \caption{Results of BMD estimation. (a) Correlation of the predicted BMD with DXA-BMD and QCT-BMD. (b) Boxplot of AE of the predicted BMD. The BMD predicted using the proposed method clearly shows a higher correlation with DXA-measured and QCT-measured BMDs and smaller absolute errors.}
    \label{fig:BMD_comp}
\end{figure}

\subsection{Results of x-ray image decomposition}
The decomposition accuracy of PSNR and DC is shown in Figure \ref{fig:epoch} (a), where significant improvement by HL was observed. The high performance of HRFormer Generator with HL in DC indicated the ability to maintain the silhouette of the decomposed structure, and the high PSNR suggested the superior capability of the quantitative decomposition compared with the same generator without HL and the baseline method. The representative cases in Figure \ref{fig1} also suggested the accurate recovery of the density distribution of PF-DRR despite the noise and variation of the overall intensity in the input x-ray image. Figure \ref{fig:epoch} (b) shows the progress of training for each backbone with and without HL, in which the robust convergence was achieved consistently using HL even with few epochs. One case was randomly chosen to track the progress during training, which is shown in Figure \ref{fig:epoch} (d) and (e). The qualitative comparison demonstrated that the target region was well-formed in the early epoch using HL, suggesting the effectiveness of HL. A summary of the experimental results for all backbones is shown in Table \ref{table:exp5}; and a detailed comparison of decomposition results between HRFormer without and with HL can be found in supplemental video.

\subsection{Results of BMD estimation}
A comparison of the BMD estimation performance between the conventional method \cite{Hsieh2021}, which uses a regression model, and the proposed HRFormer Generator with HL is shown in Figure \ref{fig:BMD_comp}. The results suggest the failure of the conventional method, which achieved an ICC of 0.361 and PCC of 0.447 under the limited dataset. In contrast, the proposed method achieved high ICC and PCC of 0.882 and 0.888, respectively, demonstrating the effectiveness of the estimation strategy of the proposed method that extracts the density distribution of the target region of the bone. Furthermore, we evaluated the prediction error in terms of T-scores. The T-scores were calculated based on the mean and standard deviation of DXA-BMD for Japanese young adult women reported in the literature (proximal femur: 0.875 $\pm$ 0.100 $g/cm^{2}$ \cite{Soen2013}). We found that the absolute error in T-score for HRFormer with HL was 0.53 $\pm$ 0.47. We additionally evaluated 13 cases whose repeated x-ray images (acquired in the standing and supine positions on the same day) were available. The coefficient of variation was 3.06\% $\pm$ 3.22\% when the best model, HRFormer with HL, was used.

\begin{table}[]
\caption{Summary of the experimental results.}
\small
\centering
\begin{tabular}{|l|cccc|ccccc|}
\hline
\multicolumn{1}{|c|}{}                         & \multicolumn{4}{c|}{Image Decomposition Accuracy}                                  & \multicolumn{5}{c|}{BMD Estimation Accuracy} \\ \cline{2-10} 
\multicolumn{1}{|c|}{}                         & mean              & mean           &                       &                       & mean           &                       &                       &                       & PCC                       \\
\multicolumn{1}{|c|}{\multirow{-3}{*}{Method}} & PSNR              & DC             & \multirow{-2}{*}{ICC} & \multirow{-2}{*}{PCC} & AE             & \multirow{-2}{*}{SEE} & \multirow{-2}{*}{ICC} & \multirow{-2}{*}{PCC} & wrt QCT                   \\ \hline
ResNet                                         & 30.688            & 0.658          & -0.278                & 0.006                 & 0.117          & 0.157                 & -0.024                & -0.208                & -0.130                    \\
+ HL                                           & \textbf{39.105}   & \textbf{0.952} & \textbf{0.866}        & \textbf{0.894}        & \textbf{0.057} & \textbf{0.074}        & \textbf{0.872}        & \textbf{0.879}        & \textbf{0.818}            \\ \hline
DAFormer                                       & 36.874            & 0.926          & 0.538                 & 0.671                 & 0.085          & 0.114                 & 0.639                 & 0.680                 & 0.632                      \\
+ HL                                           & \textbf{37.994}   & \textbf{0.945} & \textbf{0.853}        & \textbf{0.875}        & \textbf{0.057} & \textbf{0.078}        & \textbf{0.856}        & \textbf{0.865}        & \textbf{0.799}             \\ \hline
HRNet                                          & 36.996            & 0.931          & 0.369                 & 0.650                 & 0.097          & 0.127                 & 0.537                 & 0.581                 & 0.640                      \\
+HL                                            & \textbf{39.971}   & \textbf{0.958} & \textbf{0.883}        & \textbf{0.920}        & \textbf{0.057} & \textbf{0.074}        & \textbf{0.870}        & \textbf{0.878}        & \textbf{0.843}             \\ \hline
HRFormer                                       & 36.594            & 0.927          & 0.255                 & 0.495                 & 0.109          & 0.143                 & 0.313                 & 0.400                 & 0.498                      \\
+ HL                                           & \textbf{40.168}   & \textbf{0.961} & \textbf{0.910}        & \textbf{0.927}        &\textbf{0.053}  & \textbf{0.071}        & \textbf{0.882}        & \textbf{0.888}        & \textbf{0.853}             \\ \hline
\end{tabular}
%\caption{Summary of the experimental results.}
\label{table:exp5}
\end{table}

\subsection{Implementation details}
In implementing the methods for decomposing an x-ray image, we replaced the Batch Normalization and Instance Normalization with the Group Normalization \cite{Wu2018}, except for the Layer Normalization used in the Transformer. We used the structure of the Global Generator used in \cite{Wang2018} for ResNet Generator. The HRNetV2-W48 \cite{Wang2020} and HRFormer-B were used for HRNet Generator and HRFormer Generator, respectively. We set the $\lambda_{L1}$, $\lambda_{GC}$, and $\lambda_{FM}$ to 100, 1, and 10, respectively. In the stage-one training for all generators, the initial learning rate and weight decay were 2e-4 and 1e-4, respectively. In the stage-two training, the initial learning rate and weight decay were 2e-4 and 1e-8, respectively, for the ResNet Generator; 5e-6 and 1e-2, respectively, for the DAFormer Generator; 1e-4 and 1e-4, respectively, for the HRNet Generator; and 1e-4 and 1e-2, respectively, for the HRFormer Generator. We used AdamW optimizer for all decomposition methods in both stages. For data augmentation, rotation (+-25), shear(+-8), translation(+-0.3), scaling(+-0.3), and horizontal and vertical flipping were used randomly. To implement conventional method \cite{Ho2021} for BMD estimation, we followed most settings and training protocols, including the model structure, and data augmentation; however, the total number of epochs was set to 400, and we did not perform validation during training because we found that validation makes the performance worse under the limited dataset. Our implementation is available at \url{https://github.com/NAIST-ICB/BMD-GAN}.

\section{Discussion and Conclusion}
In this study, we proposed an HL framework for the image decomposition task, specifically focusing on the quantitative recovery of density distribution for small-area targets under a limited dataset. The HL reduced the demands of the dataset and had improved performance compared with conventional training. Furthermore, we experimentally compared the abilities of the generators with those of state-of-the-art backbones. With HL, all models showed significant improvement, and among them, the HRFormer Generator showed the best performance. 

We proposed a BMD estimation method, leveraging the ability of the model to decompose an x-ray image into the DRR of the proximal femur region. The proposed BMD-GAN achieved high accuracy under the limited dataset where the conventional regression model-based method failed using the same dataset. 
By training using QCT data and x-ray images, the proposed method can target the BMD of any bone within the field of view of QCT and x-ray images (unlike previous methods based on training using BMD values and x-ray images).
One limitation of this experiment was the use of QCT, which needs a special phantom in its image. However, some studies showed that BMD estimation had sufficient accuracy even using phantom-less CT data \cite{Mueller2011,Uemura2022}.
Our future work will include validation of training using phantom-less CT and the extension to other anatomical areas, such as the vertebrae and ribs. Furthermore, We plan to validate the performance of the proposed method using large-scale multi-institutional datasets.

\end{document}

%
% \title{Contribution Title\thanks{Supported by organization x.}}
\title{Supplementary Material\newline BMD-GAN: Bone mineral density estimation using x-ray image decomposition into projections of bone-segmented quantitative computed tomography using hierarchical learning}
\titlerunning{BMD-GAN}
\newcommand{\repeatthanks}{\textsuperscript{\thefootnote}}
\author{****************************************** ************************\inst{1}, ********** ******************\inst{2}}
\institute{********************************************************************** \and ************************************************** \\
\email{****************}}

\author{Yi Gu\inst{1}, Yoshito Otake\inst{1}, Keisuke Uemura\inst{2}, Mazen Soufi\inst{1},\newline Masaki Takao\inst{3}, Nobuhiko Sugano\inst{4}, and Yoshinobu Sato\inst{1}}
% index{Gu, Yi}
% index{Otake, Yoshito}
% index{Uemura, Keisuke}
% index{Soufi, Mazen}
% index{Takao, Masaki}
% index{Sugano, Nobuhiko}
% index{Sato, Yoshinobu}

%
\authorrunning{Yi Gu et al.}
\institute{Division of Information Science, Graduate School of Science and Technology,\newline Nara Institute of Science and Technology \\
\email{\{gu.yi.gu4,otake,yoshi\}@is.naist.jp}
\and Department of Orthopaedics, Osaka University Graduate School of Medicine \\
\and Department of Bone and Joint Surgery,\newline Ehime University Graduate School of Medicine \\
\and Department of Orthopaedic Medical Engineering,\newline Osaka University Graduate School of Medicine \\
}
\maketitle              % typeset the header of the contribution

\begin{figure}
    \centering
    \includegraphics[width=\textwidth]{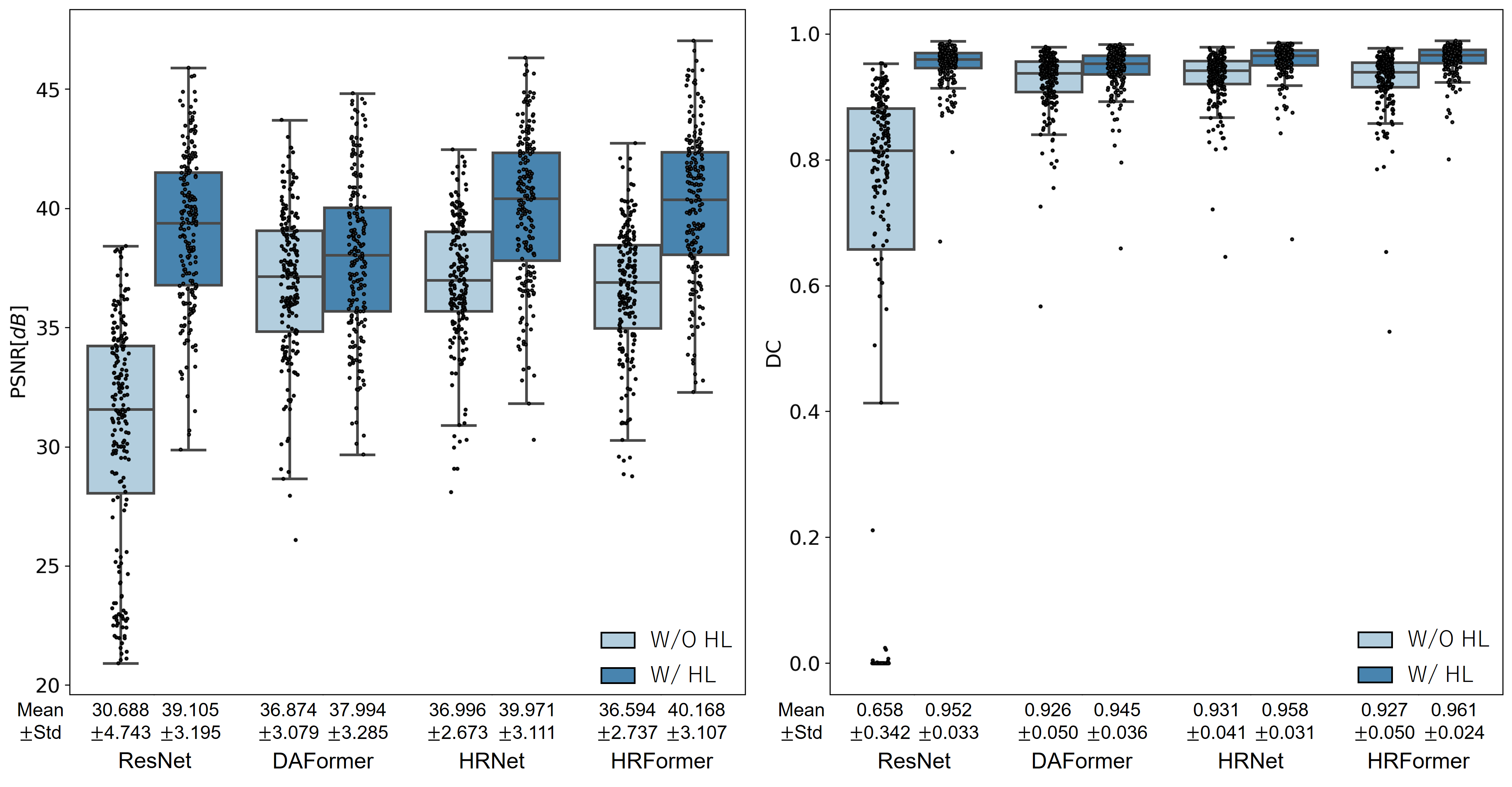}
    \caption{Quantitative evaluation results of image decomposition for different backbones w/o and w/ hierarchical learning (HL). The performance improvements by HL were observed for all the backbones, where the ResNet backbone benefited the most from HL. The lowest accuracy among the methods with HL (DAFormer w/ HL) was higher than the best accuracy among the methods without HL (DAFormer w/o HL). The comparison between HRNet and HRFormer suggests that HRNet slightly outperformed the HRFormer under no HL, while applying HL boosted both of them and made HRFormer be a better choice.}
    \label{fig:psnr_dc_box}
\end{figure}

\begin{figure}
    \centering
    \includegraphics[width=\textwidth]{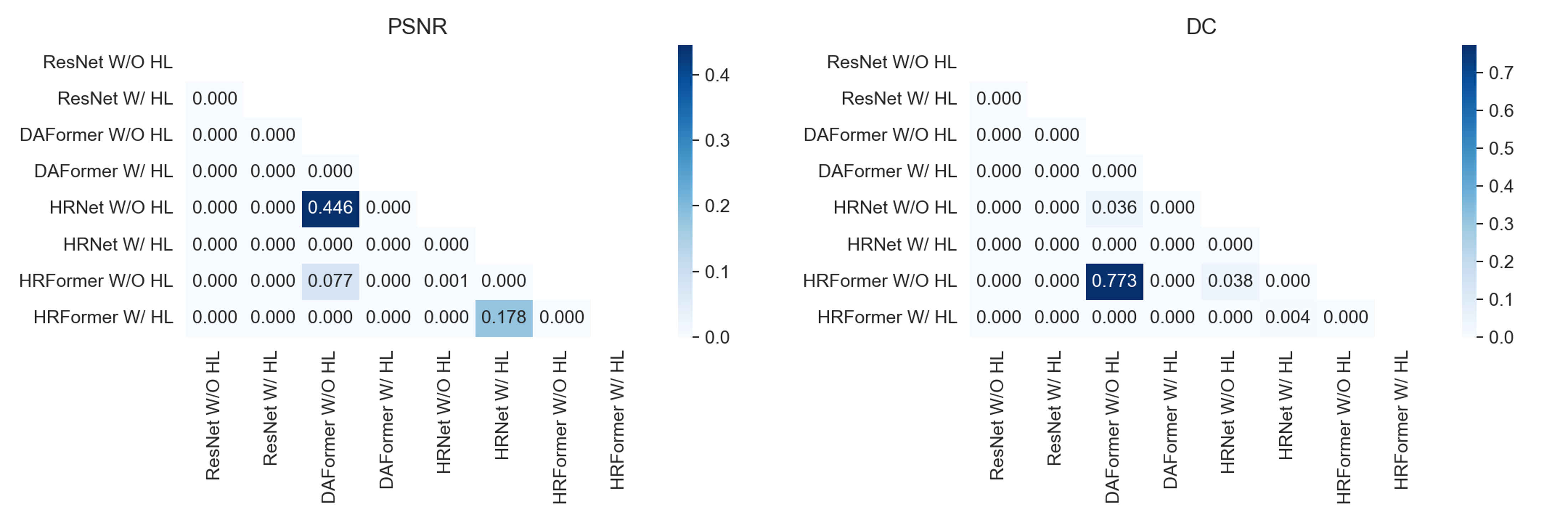}
    \caption{Results of the statistical test on PSNR and DC between methods. The $p$ values between the methods without HL and with HL were smaller than 0.001 for all backbones, indicating significant difference. The test results showed the DAformer, HRNet, and HRFormer had no significant difference when HL was not used, whereas the differences were more pronounced when HL was used.}
    \label{fig:psnr_dc_heat}
\end{figure}

\begin{figure}
    \centering
    \includegraphics[width=\textwidth]{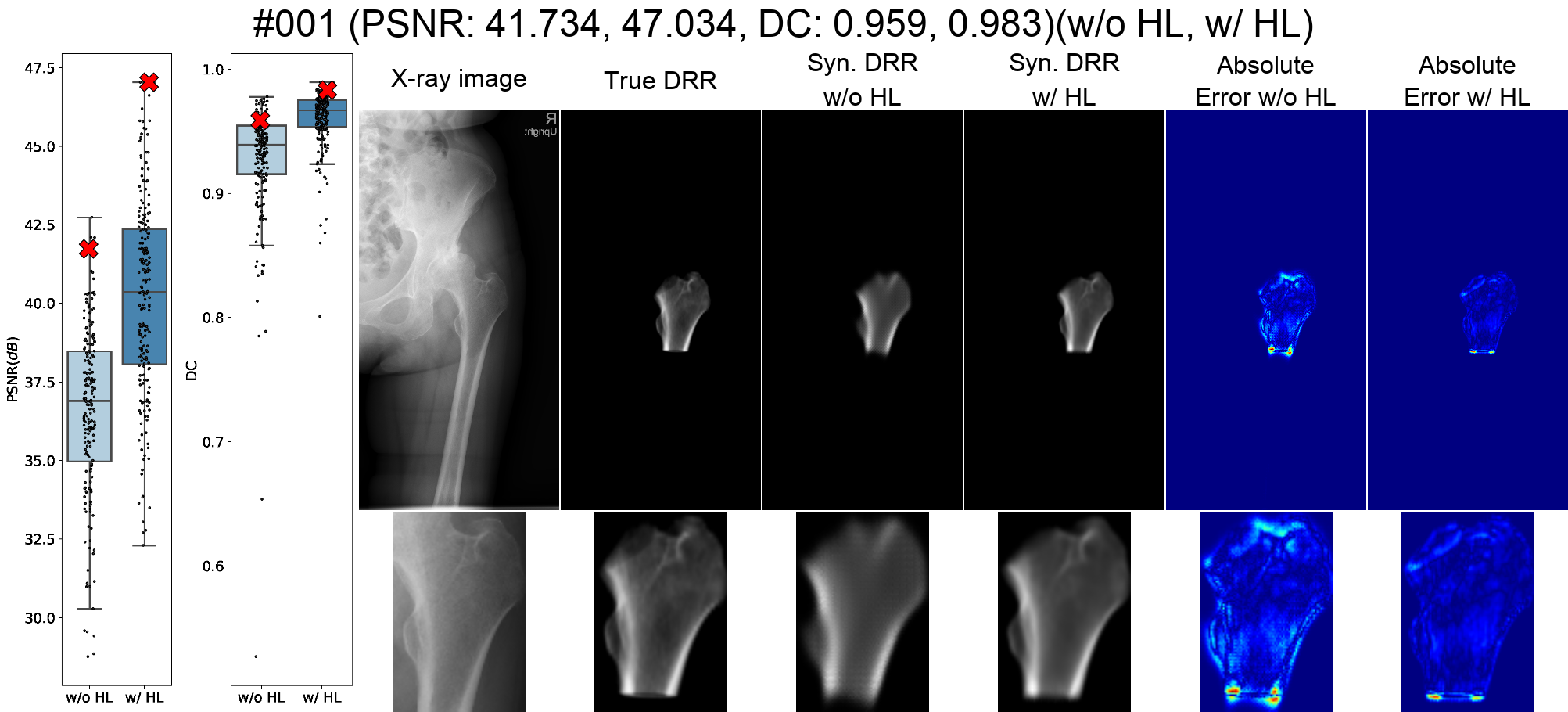}
    \caption{Screenshot of visualization video of 200-cases, visualizing decomposition results by HRFormer w/o and w/ HL in the descending order of PSNR by HRFormer w/ HL, where the more consistent shape and more detailed inner structure by applying HL were observed visually. The absolute error maps suggest the significant error is more likely produced by the upper and bottom cutting edges, which are clinically defined while applying HL made the model better in handling the cutting edges.}
    \label{fig:psnr_dc_heat}
\end{figure}

% \begin{figure}
%     \centering
%     \includegraphics[width=\textwidth]{paper1852figs/good_bad_cases.png}
%     \caption{Visualization of the x-ray image decomposition results for representative three cases. (a) X-ray images. (b) ROI images. The best results (\#92), median results (\#98), and the worst results (\#167) were shown.}
%     \label{fig:good_bad_cases}
% \end{figure}
%. \#92: 47.034 (PSNR), 0.983 (DC); \#98: 40.185, 0.939; \#167: 32.301, 0.926}